\magnification\magstep2
\centerline{\bf A quantum Coupling Conjecture}
\centerline{by}
\centerline{James G. Gilson, Mathematics QMW London}
\centerline{PIRT Conference September 2000 Imperial College London}
\centerline{(A condensed version of a paper with the same title)}

\vskip .3cm
\centerline{\bf Abstract}
\vskip 0.3cm
Following the successful prediction of an exact value for the fine structure constant later confirmed to differ numerically from the centre value of the latest experimental recommended CODATA range by $10^{-12}$, further analysis and predictions of exact values for two other quantum coupling constants, the strong and the electroweak, are given. The method employed to obtain these theoretical values depends on the conjecture that measured values of a fundamental set of quantum coupling constants approximate to exact values that can be found in a specific set $C_Q$ of numerical values which has a countable number of elements. The letter $C$ and its subscript $Q$ stand for coupling and quantum respectively. A provisional definition of this set to be refined in the body of this paper follows:
$$C_Q=\{\alpha^\prime (n_1,n_2):n_1,n_2\quad integers\}$$ where the numerical elements $\alpha ^\prime (n_1,n_2)$  are given by
$$\alpha ^\prime (n_1,n_2) = \alpha (n_1,n_2)= n_2\cos (\pi/n_1)\tan (\pi/(n_1 \times n_2))/\pi$$
together with elements of the form
$$\alpha^\prime (n_1,n_2) = (\alpha (n_1,n_2)+ \alpha (n_1 + 1,n_2))/2.$$
The inclusion of the arithmetic mean values is to take into account measurements that take place on an energy boundary between two possible consecutive theoretical values given by the function $\alpha (n_1,n_2)$ and so are unable to discriminate between them.
Questions of how this conjecture might be validated from the theoretical and measurement point of view and the identification of those elements of $C_Q$ which have values of definite physical significance constitute the subject matter of this paper.
\vskip 0.3cm

\break
\centerline {\bf Origins}
\vskip 0.3cm
The basic formula for $\alpha$ is given in the abstract above. This formula for the numerical value of $\alpha$ originated in a stochastic theory of a {\it mass\/} polarized vacuum
alternative to Sch\"odinger theory. Circular orbit representation for states in a 6-space are the basis elements in the alternative  theoretical construction in which negative mass plays an {\it equal\/} role with positive mass in strong contrast with the situation in the orthodox theory.  A derivation of the basic formula was subsequently obtained from the {\it orthodox\/} theory and a refinement factor involving the second quantum number $n_2$ was introduced. This additional quantum number will now be shown to play a vital part in extending the theory to evaluating the other quantum
coupling constants.
\vskip 0.3cm
\centerline{\bf CODATA's 1986 $\alpha$ value}
\vskip 0.3cm
The CODATA  experimental reliability range for the value of $\alpha$ changed in 1999 to {\it outside\/} its recommended value range in 1986. This dramatic change in the assessed value for $\alpha$ that has been assumed correct for the last thirteen years must through doubt on the current values that are being claimed for the measured values of the other coupling constants which are so very much less well understood theoretically. The value of measurement comes from clear theoretical understanding of what is being measured. However, outside QED theory is not strong. The current measurement information is very varied coming from about a dozen different sources all with differing theoretical interpretations, omissions and additions of imagined theoretical possibilities. I have opted to stick with what is called the modified minimal subtraction scheme $\overline{MS}$, though with little confidence of its exactitude. The scheme I am about to propose to predict from theory the value of 
the other coupling constants must consequently be regarded as provisional in form and in the predictions it generates as it depend {\it strongly\/} on having correct physical input concerning reliability {\it ranges\/}. CODATA's erroneous placing of the range for measured $\alpha$ values led me into choosing the wrong value for the second parameter in my $\alpha$ formula. I chose $n_b=n_2=25$ as that gave a theoretical value close to their central recommended value. In the last few years, for various reasons, I came to realize they had the wrong range and the correct value for the second parameter should be $n_2=29$. The value $29$ has a substantial significance for evaluating the other coupling constants.
   
\break  
\centerline{\bf A New Relativity Quantum Link}
\vskip 0.5cm

Relativity is usually regarded as a not-quantized theory and much effort has gone into what is seen as the need to give it a quantization structure in order to make it join with or become compatible with quantum mechanics. However, relativity does have a fundamental quantization in the definite and unique fixed value of the velocity of light in the vacuum. This is not usually thought of as quantization presumably because velocity is not one of quantum mechanics {\it dynamical\/} variables. However, here I think an important connection between relativity and quantum mechanics has been overlooked. The work on the fundamental coupling constants that I have been engaged on for some years has thrown up a {\it new\/} connection between relativity and quantum mechanics that could be called {\it the triangle link}. This was briefly discussed in my paper for our last conference though not called by that name. It takes the form of a connection between quantum's projection quantization princ
iple and {\it velocity\/} quantization in relativity. As was discussed in the last conference paper there appears a really comprehensive relativistic quantization in the relativistic analysis of the motions of electrons in first Bohr orbits. The triangle link can be briefly described as follows. When a right angled triangle rotates about one of its non-$\pi /2$ vertices fixed in space the non-radial side will be subject to a length contraction whereas the hypotenuse will, in some quantum situations, have a quantized projection on to the other radial side. This implies by Pythagoras theorem that there will be a related quantized velocity associated with the non-radial side's relativity contracted length. The quantized velocities involved in this connection are all sub-light velocity but there is one seemingly important special case $v=137\alpha c$ just below the $c$ maximum value.    
\vskip 0.5cm

\centerline {\bf The complementarity theorem }
\vskip 0.5cm

It is obvious from inspecting the wave capture diagram that the relativistic wave capture process capture involves the geometry of polygons. In my first introduction of this wave capture idea a $137$ sided polygon was introduced together with its side number being a new type of $\it physical\/$ eigenvalue. In fact, the side number was essentially the result of an angular quantization induced by the sub-angle $\pi /137$. As the work evolved the more accurate result for the value of $\alpha$ emerged from the more refined angular quantization $\pi /(29\times 137)$ involving the number $29$. It soon became apparent that the value of the fine structure constant $\alpha (137,29)$ was a characteristic value associated with representations of one factor of the product group of rotations associated with a $29\times 137$ sided polygon. I then naturally considered the question: what does the equivalent characteristic $\alpha (29,137)$ of the other factor of the $29\times 137$ rotation group 
represent. To my very great surprise the answer turned out to be $\alpha (29,137)\approx \alpha _g$. I have put $\approx$ here because the value of $\alpha _g$ is not known very accurately.  However, the agreement seems too good to be just an accident. Thus in the following work, I have made the perhaps over optimistic assumption that if the value of $\alpha$ at some energy is $\alpha (n_1,n_2)$ then at the same energy the value of $\alpha _g $ is  $\alpha (n_2,n_1)$. This assumption I call the complementarity theorem for the electromagnet and electro-weak couplings. The factor groups can be seen to be associated with the {\it supplementary\/} quantized angles $\pi/137$ and $\pi/29$ at low energies. 
\vskip 0.5cm

\break

\centerline {\bf The set $C_Q$}
\vskip 0.5cm
Provisional identification of the elements of $C_Q$
that correspond to $\alpha$, $\alpha _g$ and $\alpha _s$ at some
approximately known energy values are given in the following table.
\vskip 0.5cm
$$*\alpha _s(m_\tau)=0.334\pm 0.022\to\alpha(2,1)=1/\pi =0.318309886183...$$
$$*\alpha _s(m_Z)  =0.118\pm 0.003\to\alpha(8,1)=0.12811...\  or\ \alpha(8,2)=0.11699...$$
$$\alpha _s(m_U) =\alpha[42,42]=0.023742972654...$$

$$\alpha _g(m_0)=\alpha[29,137]=0.034280626357...$$
$$\alpha _g(m_\tau)\approx 0.034\quad\quad\quad\alpha(30,133)=0.033150736696...$$
$$*\alpha _g(m_Z)=\alpha (m_Z)/0.23117\approx 0.0338129\to\alpha(31,128)=0.03209256552...$$
$$\alpha _g(m_U) =\alpha[42,42] =0.023742972654...$$

$$\alpha (m_0) =\alpha (137,29)=0.0072973525318...$$
$$*\alpha (m_\tau)^{-1} =133.531\pm 0.026\to{{\alpha (133,30)+\alpha (134,30)}\over{2}}=0.007...\approx{1\over{133.5}}$$
$$*\alpha (m_Z)^{-1} =127.934\pm 0.027\to\alpha (128,31)=0.00781...\approx {1\over{128.039}}$$
$$\alpha (m_U) =\alpha[42,42] =0.023742972654...$$

\vskip 0.5cm
\centerline{\bf 5. The Calculation}
\vskip 0.5cm
The calculation is aimed at finding the three running coupling constants $\alpha _s(x),\alpha _g(x)$, and $\alpha _f(x)=\alpha (x)$ as a function of $xmass(Q)=log_{10}(Q)$. $xmass$ will be abbreviated to plain $x$ most of the time, $xmass$ is used rather than energy because of the extremely large energy range involved, 0 to $\approx 10^{16}$ Gev. Clearly, the structure is easily re-expressed in terms of energy $Q$ if that should become necessary at some stage. The overall scheme of the three running coupling constants taken together is to be constructed so that the relations between them are such that they mutually converge at the value $x_U=16$. Values of this order are suggested from grand unified theory. These objectives require the finding of six integer values functions of $x$, two for each of the coupling constants,
$$s(x),s_b(x),g(x),g_b(x),f(x),f_b(x).$$
However, some simplification arises from the complementarity theorem for $\alpha_g$ and $\alpha _f$ which gives two relations between these six functions, $$g(x)=f_b(x),g_b(x)=f(x).$$
These functions select out the integer parameters that are to be put into the basic formula $\alpha (n_1,n_2)$ for a coupling constant so as to define a definite path through the set $C_Q$ determined by the $x$ parameter representation for energy. Thus the three running coupling constants assume the forms
$$\alpha_s(x)=\alpha(s(x),s_b(x)),$$
$$\alpha_g(x)=\alpha(f_b(x),f(x))$$ and
$$\alpha_f(x)=\alpha(f(x),f_b(x)).$$
The problem now becomes to find the paths through the set $C_Q$ as the energy related parameter $x$ ranges from some low or negative value up to its grand unified value $16$ or even beyond that value. The final set of values assumed by the s and f functions are to be integers so as to pick up the true members of the set $C_Q$. That is to say they are step functions with many steps. However, such functions can clearly have smooth continuous approximations that are mathematically much easier to find and work with than exact step forms. We can use simple polynomials of the variable $x$ for these continuous smooth functions and find them by solving the energy value identification equations for their coefficients. Once we find {\it good\/} smooth approximate versions which thread the elements of $C_Q$ we can obtain the step versions by taking the integer valued part at all values of $x$ along the smooth paths. There is considerable freedom in setting up the equations for the smooth app
roximations because in principle they can, at the energy identification points, have any value from the actual final integer sought to just below the next higher integer. This freedom allows us to set up step widths on which the function value is constant. However, the physical measurement information is so very sparse very little use can be made of this possibility at the present time. However, as more physical information becomes available improvements are likely to be made using this freedom. Thus inspecting table 1, it is seen that a set of equations such as follows for the given continuous functions can be solved to find their coefficients. This is a first attempt at the problem given the information in table 1 only, and using polynomials of order just sufficient to take up the information therein. Two options $n_{sb1}$ and $n_{sb2}$ are given for the second parameter of the strong coupling constant to take into account the uncertainty caused by the two-option identification in table 1 for $\alpha _s(m_Z)$. Hopefully, this question will be resolved by future measurement.

\break
The necessary five continuous functions are taken to be,

$$n_s(x)=a_s x^2 + b_s x + c_s$$
$$n_{sb1}(x)=a_{sb1} x^2 + b_{sb1} x + c_{sb1}$$
$$n_{sb2}(x)=a_{sb2} x^2 + b_{sb2} x + c_{sb2}$$
$$n_g(x)=a_g x^3 + b_g x^2 + c_g x + d_g$$
$$n_f(x)=a_f x^3 + b_f x^2 + c_f x + d_f.$$

Seventeen simple equations used to determine the coefficients are,

$$n_s(x_\tau)=2,\ n_s(x_Z)=8.001,\ n_s(x_U)=42$$
$$n_{sb1}(x_\tau)=1,\ n_{sb1}(x_Z)=1.01,\ n_{sb1}(x_U)=42$$
$$n_{sb2}(x_\tau)=1,\ n_{sb2}(x_Z)=2.01,\ n_{sb2}(x_U)=42$$

$$n_g(x_0)=29,\ n_g(x_\tau)=30,\ n_g(x_Z)=31,\ n_g(x_U)=42$$

$$n_f(x_0)=137,\ n_f(x_\tau)=134.000001,\ n_f(x_Z)=128.0001,\ n_f(x_U)=42.$$

The measured values of the xmass parameter used here as inputs are $$xmass(m_0)=log_{10}(0.51099906/1000)\approx -3.29158$$
$$xmass(m_\tau)=log_{10}(1777.0/1000)\approx 0.24969$$
$$mass(m_Z)=log_{10}(91.187)\approx 1.95993$$ in terms of the corresponding particle rest masses. $xmass(m_0)$ is the electron's energy index.

The integer valued functions to be used in the running functions against $x$ for the coupling constants are now given by taking the greatest integer value less than or equal to the corresponding continuous function at each $x$ position. This operation is sometimes called {\it Floor\/}.
$$s(x)=Floor(n_s(x)),\ g(x)=Floor(n_g(x)),\ f(x)=Floor(n_f(x)),\ etc...$$
The integral functions of $x$ can then be used directly as parameters in the formula $\alpha (n_1,n_2)$ to generate the three running coupling constants.

\break
The provisional set of predicted values for $ \alpha _{s1}, \alpha _{s2} , \alpha _g , \alpha $ respectively  against integral values of the
energy related parameter $xmas$ at the first entry far left of table below. Two options are given for the strong coupling constant at the second and third entries.

\vskip 0.3cm
$$\vbox{\halign{#&\ #&\ #&\ #&\ #\cr
$\ x$&$\alpha _{s1}\quad\quad\quad$&$\alpha _{s2}\quad\quad\quad $&$\alpha _g\quad\quad\quad $&$\alpha _f\quad\quad\quad $\cr
$-3$ & $0.318309886184$ & $0.318309886184$ & $0.03428062635$ &
 $0.007297352532$\cr
$-2$ & $0.318309886184$ & $0.318309886184$ & $0.03428062635$ &
 $0.007297352532$\cr
$-1$ & $0.318309886184$ & $0.318309886184$ & $0.03428062646$ & $0.007350981026$\cr
$00$ & $0.318309886184$ & $0.318309886184$ & $0.03428062668$ & $0.007460637335$\cr
$01$ & $0.225079079039$ & $0.225079079039$ & $0.03315073690$ & $0.007631394409$\cr
$02$ & $0.121811919801$ & $0.116992293559$ & $0.03209256563$ & $0.007871608420$\cr
$03$ & $0.089678258102$ & $0.087491104356$ & $0.03209256608$ & $0.008127431400$\cr
$04$ & $0.069931404971$ & $0.069710854585$ & $0.03109953000$ & $0.008543929592$\cr
$05$ & $0.057895194158$ & $0.057848290373$ & $0.03016582310$ & $0.008925061313$\cr
$06$ & $0.049400670268$ & $0.049392707894$ & $0.02928630686$ & $0.009519549431$\cr
$07$ & $0.043078769656$ & $0.043076609395$ & $0.02845641619$ & $0.010095927437$\cr
$08$ & $0.039687167158$ & $0.039686038750$ & $0.02845641742$ & $0.010863231950$\cr
$09$ & $0.035490756013$ & $0.035490481618$ & $0.02767208466$ & $0.011756675426$\cr
$10$ & $0.033151268431$ & $0.033151149158$ & $0.02692967248$ & $0.012810120421$\cr
$11$ & $0.030166068278$ & $0.030166022483$ & $0.02622591956$ & $0.014070727859$\cr
$12$ & $0.028456566299$ & $0.028456541078$ & $0.02555789319$ & $0.015606187244$\cr
$13$ & $0.026929750625$ & $0.026929744397$ & $0.02492294857$ & $0.017216103865$\cr
$14$ & $0.026225969886$ & $0.026225966061$ & $0.02492295229$ & $0.019195698331$\cr
$15$ & $0.024922972884$ & $0.024922970776$ & $0.02431870057$ & $0.021688471563$\cr
$16$ & $0.023742972654$ & $0.023742972654$ & $0.02374297265$ & $0.023742972654$\cr
$17$ & $0.023193792720$ & $0.023193792720$ & $0.02374297989$ & $0.026929698539$\cr}}$$

\break
\centerline {\bf Conclusions}
\vskip 0.5cm

This attempt to find the exact values of the quantum coupling constants stands or falls according to whether the initial conjecture is correct or not. If the conjecture is not correct then the above calculation is useless. However, if the conjecture is correct the calculation could fail to deliver the correct values because the physical input information is inadequate. In this case, judging by the rate at which the measurement process has developed in the past decade it is possible that we shall have to wait another decade for the necessary extra decimal places to be found. I have taken the optimistic view that the conjecture is correct and that present measurement information is sufficiently near to reality for this first attempt at the calculation to stand a reasonably good chance of succeeding. Any way, if the conjecture is correct, it gives a certain indication of a collection of exact values that will still only be obtained approximately in a measurement process. The conjectu
re is a very strong restriction on the possible approximate values that may be obtained by measurement. However, the uncertainty lies in the association of an energy with any measured value as that depends on the positions of the step edges which at the present time can only be obtained from measurement or by a clever choice of the shape of the continuous curves that are used as an auxiliary aid in the calculation. 
\vskip 0.5cm
\centerline {Acknowledgements}
\vskip 0.5cm
I am very grateful to Professor C.W.Kilmister for many discussions and much guidance and to Professor Wolfgang Rindler for inspirational lectures on relativity many years ago.

\end